\documentclass[manuscript]{aastex}

\shorttitle{Galactic Reddening Predictions}
\shortauthors{Burstein}

\begin{document}

\title{Line-of-Sight Reddening Predictions: Zero Points, Accuracies,
the Interstellar Medium, and the Stellar Populations of Elliptical Galaxies}

\author{David Burstein\altaffilmark{1}}

\altaffiltext{1}{Department of Physics and Astronomy, Box 871504, Arizona
    State University, Tempe, AZ  85287-1504}

\begin{abstract}

Revised (B-V)$_0$-Mg$_2$ data for 402 elliptical galaxies are
presented which are used to test reddening predictions.  These
reddening predictions can also tell us both what the intrinsic errors
are in this relationship among gE galaxy stellar populations, as well
as details of nearby structure in the interstellar medium (ISM) of our
Galaxy and of the intrinsic errors in reddening predictions. Using
least-squares fits, the explicit $1\sigma$ errors in the
Burstein-Heiles (BH) and the Schlegel et al. (IR) predicted reddenings
are calculated, as well as the $1\sigma$ observational error in the
(B-V)$_0$-Mg$_2$ for gE galaxies.  It is found that, in directions
with E(B-V)$<$0.100 mag (where most of these galaxies lie), $1\sigma$
errors in the IR reddening predictions are 0.006 to 0.009 in E(B-V)
mag, those for BH reddening prediction are 0.011 mag, and the
$1\sigma$ agreement between the two reddening predictions is 0.007
mag.  IR predictions have an accuracy of 0.010-0.011 mag in directions
with E(B-V)$\ge$0.100 mag, significantly better than those of the BH
predictions (0.024-0.025). Both methods yield good evidence that
gas-to-dust variations that vary by a factor of 3, both high and low,
exist along many lines-of-sight in our Galaxy. Both methods also
predict many directions with E(B-V)$<0.015$ mag, despite the
difference in zero point that each has assumed. The $\sim$0.02 higher
reddening zero point in E(B-V) previously determined by Schlegel et
al.  is confirmed, primarily at the Galactic poles. Independent
evidence of reddening at the North Galactic pole is reviewed, with the
conclusion that there still exists directions at the NGP that have
E(B-V)$<<0.01$.  Two lines of evidence suggest that IR reddenings are
overpredicted in directions with high gas-to-dust ratios. As high
gas-to-dust directions in the ISM also include the Galactic poles,
this overprediction is the likely cause of the E(B-V)$\sim$0.02 mag
larger IR reddening zero point relative to that of BH.

\end{abstract}


\section{Introduction}

Maps of line-of-sight Galactic reddening \citep{bh78b} ushered in a
new era in astronomy in which reddenings could be reliably predicted
along arbitrary lines-of-sight.  These maps supplanted the cosecant
laws previously used to make such predictions \citep[e.g.,][]{deV76}.
The more recent method for predicting line-of-sight reddening
\citep{sfd98} is also differentially accurate in its predictions, but
assumes a zero point $\sim 0.02$ mag higher in E(B-V) than does the
Burstein \& Heiles method.  Cursory examination of the reddening
predictions listed for the Burstein \& Heiles method and that of
Schlegel et al.  on the NASA Extragalactic Database (NED) website
(nedwww.ipac.caltech.edu) shows that this difference in zero point
produces a large difference in predicted B mag extinction ($4.315
\times 0.02 = 0.086$ mag), and belies the actual good differential
agreement between the two reddening predictions found in the present
paper.  Yet it is the 20-year experience of this author that very few
astronomers take cognizance of this zero point issue.  Schlegel et
al. further claim that their method is significantly more
differentially accurate than the Burstein \& Heiles method (see the
abstract and Sec. 5 of their paper).

\S 2 of the present paper presents the (B-V)$_0$-Mg$_2$ data for gE
galaxies used for the reddening calibration tests in both this paper
and in \citet{sfd98}.  These data are a significant advance in
calibrating {\it differential} reddening predictions over what was
available from globular clusters and RR Lyrae stars in \citet{bh78b}.
In \S 2, via a set of four coupled least-squares equations, $1\sigma$
errors are explicitly derived in a least-squares sense for the
Burstein \& Heiles and the Schlegel et al. reddening predictions, as
well as for the (B-V)$_0$-Mg$_2$ relation for gE galaxies.  \S 3
discusses various related issues, including: reddening at the North
Galactic Pole (NGP); HI gas-to-dust variations in the interstellar
medium (ISM); minimum line-of-sight reddenings; the stellar
populations of gE galaxies; and systematic issues surrounding
reddening predictions.  The case in made in \S 4 that the zero point
difference between the Burstein \& Heiles reddening predictions and
the Schlegel et al.  reddening predictions is due to Schlegel et al.
overpredicting reddenings in directions where gas-to-dust ratios are
high. The major findings of this paper are summarized in \S 5.

\section{Accuracy of the Reddening Predictions}

\subsection{The Reddening Calibration Data: Elliptical Galaxy
Stellar Populations}

The reddening prediction method developed by Burstein \& Heiles
\citep[][hereafter termed the ``BH'' method]{bh78a,bh78b,bh82,bh84}
uses the empirical fact that galaxy counts decrease with increasing
extinction to monitor the varying HI column density-to-dust ratio that
was discovered in \citet{bh78b}.  The availability of maps of infrared
emission in the sky in the 1980's and 1990's made it feasible to make
new line-of-sight predictions of reddening using observations of
emission by dust \citep[][hereafter termed the ``IR'' method]{sfd98},
even if it is still not known what portion of the dust in the ISM
produces the reddening that is seen in the optical
\citep[cf.][]{lg97}.  One must assume a relation between extinction
and E(B-V) to calculate extinction from these methods (e.g., a value
of 4.315 is assumed by Schlegel et al. to convert E(B-V) to $\rm A_B$;
BH used 4.0 for this conversion for reasons discussed in
\citet{bh78b}).

Neither set of reddening prediction methods directly measures the dust
that is doing the line-of-sight reddening.  Rather, each must assume a
relationship between the measured quantity and the line-of-sight
reddening. Both are therefore ``differential'' predictions of
line-of-sight {\it reddening} which need to be calibrated with
external data. \citet{bh78b} calibrated their method using globular
cluster and RR Lyrae star reddenings.  Since that time, the integrated
stellar population properties of giant elliptical (gE) galaxies have
been found to be so uniform that they can be used for accurate
differential reddening measurements \citep{dbetal88,smfetal89}.

At their request, this author provided Schlegel et al. with a set of
Mg$_2$ line strengths and (B-V) colors for gE galaxies taken from the
7 Samurai \citep{smfetal89} database.  The database this author
provided Schlegel et al. differs from that published in
\citet{smfetal89} in one important respect. The
Mg$_2$-velocity-dispersion-(B-V) relationships \citep{bbf93} among gE
galaxies were used to put each subset of the 7 Samurai Mg$_2$ data on
a consistent system.\footnote{Published Mg$_2$ values
\citep[cf.][]{rldetal87} for several subsets of the 7 Samurai data
have been modified: -0.012 mag added to LCO H1 data, 0.014 mag to LCO
H2, +0.011 mag to LCO 2D, and +0.020 mag to AAT1.  (The LCO H1, LCO
H2, LCO 2D and AAT1 subsets refer to the telescopes and observing runs
used by members of the 7 Samurai to obtain spectroscopic data for
these galaxies. The LCO data was taken at the Las Campanas
Observatory, and the AAT data taken at the Anglo-Australian Telescope.
The labels H1, H2, and 2D refer to the different spectrographic
configurations used for the LCO spectra; see Davies et al. for
details.)}  The corrected Mg$_2$ measures for each data subset are
then added back together with the rest of the 7 Samurai Mg$_2$ data,
and new average values of Mg$_2$ for affected galaxies are calculated.
Since this effectively produces a new set of data, the full set of
these Mg$_2$ data are being made available with this paper
\citep[cf.,][where a portion of these data were also
presented]{hudson99}.

Table~1 gives the first 10 lines of the reddening-related data used in
this paper for elliptical galaxies; the full list for 402 gE galaxies
is available via anonymous ftp from samuri.la.asu.edu/pub/burstein/bvmg2
(bvmg2\_Table\_1\_burstein03.lst).  13 gE galaxies do not have direct
BH or IR reddenings (see the notes to Table~1 for details). A further
11 gE galaxies in the region $\rm 230^\circ < l < 310^\circ$, $\rm
-20^\circ < b < 20^\circ$ were found in \citet{dlbetal88} and
\citet{dbetal87} to have HI-only BH reddening predictions that are a
factor of two too high (These galaxies are given in the notes to
Table~1 and are noted in the ASCII version of this table by an
asterisk in their names.)  Burstein et al.  predicted reddenings for
these 11 galaxies use the (B-V)$_0$-Mg$_2$ relation, so 7 Samurai
reddenings cannot be used for these galaxies. Rather, the values of
E(B-V) for these 11 galaxies predicted from the BH method have been
uniformly halved to go with the higher gas-to-dust ratio found by the
7 Samurai. (The reddenings for these 11 galaxies are separately
discussed below.) It may be a bit confusing, but it should be noted
that an additional 11 galaxies within a smaller region (see notes to
Table~1) did {\it not} need to have their reddenings so-corrected in
the 7 Samurai database.  These other 11 galaxies include several which
are as reddened as the galaxies that {\it do} require gas-to-dust
corrections.  Evidently, this southern region has a mixture of regions
of high gas-to-dust ratios and regions of normal gas-to-dust ratios.

Twenty-four (24) galaxies are thus objectively excluded from the
least-squares error-determination analysis of this paper. This yields
378 gE galaxies with direct BH {\it and} direct IR reddening
predictions.  To make sure that the reddenings used for the IR method
are those in the public domain on the NASA Extragalactic Database
(NED) website, the NED group kindly provided this author with the IR
reddenings for the galaxies in this sample.\footnote{The NASA/IPAC
Extragalactic Database (NED) is operated by the Jet Propulsion
Laboratory, California Institute of Technology, under contract with
the National Aeronautics and Space Administration.}

\subsection{Observational and Intrinsic Accuracies of Reddening Methods}

The 378 gE galaxies are divided into three samples: those with
reddenings E(B-V)$\le$0.100 averaged from both methods (the ``low''
reddening sample); those with E(B-V)$>0.100$ (the ``high'' reddening
sample), and the full sample taken as a whole.  (The rationale for
separating galaxies into low/high reddening directions is given
below).  No separation is made at the dividing line for the BH method
(declination $ = -23.5^\circ$), south of which the BH predictions
change their methodology from use of both galaxy counts and HI column
densities to use of only HI column densities.  \citet{sfd98} did not
find a difference in these two regions, save the $l = 230-310^\circ$
region identified above.  Note that only HI values are also used at
high Galactic latitudes ($\rm b > \mid \! 65^\circ \!  \mid$) in the
BH method, as galaxy counts become correlated in real space above
those latitudes; \citep[cf., discussion in][]{bh78b}.

IR and BH reddening predictions first must be placed on the same zero
point scale. This can be done in two ways, using all 378 galaxies: a)
Zeroing IR predictions to BH predictions yields a {\it median} offset
of 0.016 mag in E(B-V), which si then subtracted from the IR
predictions; b) Zeroing BH predictions to IR predictions requires
adding a median offset of 0.019 mag in E(B-V) to the BH predictions.
Both zero point offsets are consistent with the E(B-V) = 0.02 zero
point difference between IR and BH methods found by Schlegel et
al. (estimated only in in directions with $\rm b > \mid \! 45^\circ \!
\mid$; i.e.  the Galactic poles.)

The slight difference (0.003 mag) in zero point difference calculated
these two ways is because negative BH reddenings are routinely given
for the BH method, owing to the existence of HI gas with no measured
reddening. BH acknowledged from the outset that the their zero point
could be in error by 0.01 mag in E(B-V) \citep[cf.][]{bh82}. While
negative values of reddening are explicitly calculated in the BH
method, these predicted reddenings are set to zero if they are used as
reddenings \citep[cf.][]{bh84}.  Thus, if a different reddening zero
point was established, adjustment of the BH reddening predictions
would then only require addition of the new zero point (as used
here). In contrast, the IR method defines its zero point as formally
0.000. Hence, adjusting BH zero point to IR zero point results in some
negative BH reddenings becoming positive, but adjusting IR zero point
to BH zero point results in some IR reddenings becoming negative (and
are therefore set to 0.000 for reddening calculations).  There are
351/27 low/high reddening galaxies when the 0.016 mag zero point
difference is subtracted from the IR method, and 344/34 low/high
reddening galaxies when the 0.019 mag zero point difference is added
to the BH method.  For each zero point used, averages of the two
reddening predictions (i.e, BH+IR-0.016 and IR+BH+0.019) are also
calculated.

The (B-V)$_0$-Mg$_2$ relationship for gE galaxies are fit via two
equations: (B-V)$_0$ = 1.12 Mg$_2$ + constant and (B-V)$_0$ = 0.942
Mg$_2$ + constant.  The 1.12 slope comes from \citet{bbf93}, who show
that the (B-V)$_0$-Mg$_2$ relationship extends over a much wider range
of color than is given for just the 7 Samurai galaxies.  The 0.942
slope is derived from only the gE data used here.  Use of either
slope, with the appropriate choice of constant, yields very similar
results for the gE dataset.  Constants for these equations are chosen
to yield zero {\it median} values of residuals for the 378 galaxies
sample: for the 1.12 slope, they are 0.600, 0.601, 0.601, 0.618,
0.617, and 0.618 respectively for the BH, IR and Average values
(respectively), with the BH+0.019 \& IR values listed first, and the
BH \& IR-0.016 values listed second.  Analogously, they are 0.654,
0.654, 0.655, 0.672, 0.669, 0.670 for the 0.942 slope.

It is assumed that all deviations from the (B-V)$_0$-Mg$_2$ 
relation are due to a combination of random observational 
and cosmic errors for the reddening predictions and in the
(B-V)$_0$-Mg$_2$ relationship for gE galaxies.  Handling the data
in the manner described above gives 12 ways of looking at these data:
two different zero points, two different slopes of the Mg$_2$-(B-V)$_0$ 
equation, and three subsamples: low, high and ``all'.

For each of these 12 ways, there are four observationally-independent
$1\sigma$ errors we can calculate i) the error of BH reddening
predictions on the (B-V)$_0$-Mg$_2$ relationship ($\rm
\sigma(BH:BV)$); ii) the error of the IR reddening predictions on the
(B-V)$_0$-Mg$_2$ relationship ($\rm \sigma(IR:BV)$); iii) the error of
the average (BH+IR) reddening prediction on (B-V)$_0$-Mg$_2$
relationship ($\rm \sigma(Ave:BV)$); iv) and the $1\sigma$ error for
the the agreement in reddening predictions of BH to IR ($\rm
\sigma(BH:IR)$).  (No individual weighting is given each galaxy, as
the 7 Samurai cited only average errors for these data.)  In turn,
each observable error is itself the quadratic sum of the intrinsic
errors for each data set: $\rm \sigma(BH)$, $\rm \sigma(IR)$, $\rm
\sigma(Ave)$, and $\rm \sigma(BV+Mg_2)$.  This gives us four equations
in four unknowns which relate the observable errors to these intrinsic
errors:

$${\rm \sigma(BH:BV)}^2 = {\rm \sigma(BH)}^2 + 
{\rm \sigma(BV+Mg_2)}^2 ,$$ 

$${\rm \sigma(IR:BV)}^2 = {\rm \sigma(IR)}^2 + {\rm \sigma(BV+Mg_2)}^2 ,$$ 

$${\rm \sigma(Ave:BV)}^2 = {\rm \sigma(Ave)}^2 + {\rm
\sigma(BV+Mg_2)}^2 ,$$  

\centerline{and}

$${\rm \sigma(BH:IR)}^2 = {\rm \sigma(BH)}^2 + {\rm \sigma(IR)}^2 .$$ 

Intrinsic errors are determined via a least-squares analysis.
Table~2 gives the results of this least-squares analysis, two lines
for each of the 12 ways of looking at these data (see the table 
notes for an explanation to Table~2).  Figures~1, 2 and 3
show results from using the 1.12 slope for the (B-V)$_0$-Mg$_2$
relation, as results from the 0.942 slope are virtually
indistinguishable from these data (cf. Table 2).  Figures~1a-f are 
histograms of residuals from the (B-V)$_0$-Mg$_2$ relation for the
378 gE galaxies, divided into low and high reddening subsets.
Figures~1a-c are the residuals from the (B-V)$_0$-Mg$_2$ relation for
gE galaxies using (a) the IR-0.016 values, (b) the original BH values,
and (c) the average of these two reddening predictions. Figures~1d-f
are analogous to Figures~1a-c, but calculated using (d) the original
IR values, (e) the BH+0.019 values, and (f) the average from these
values.

Figures~2a-f (parallel to Figures~1) give the scatter in these
reddening predictions versus the average value of reddening for both
samples of the data for each kinds of zero points these gE galaxies,
now including the 11 high gas-to-dust galaxies in the $230^\circ < l <
310^\circ$, $-20^\circ < b < 20^\circ$ region that were not included
in the least-squares analysis. Different symbols are used to denote
these 11 galaxies, as well as the 8 galaxies in the Perseus cluster 
(the most-reddened galaxies in this sample that are all in one
direction), as these galaxies are discussed separately below.
Figures~3a,b show the histograms of the scatter in the BH and the IR
reddening predictions for these gE galaxies, again with the two
subsets of highly-reddened samples distinguished; Figure~3a for 
BH vs. IR-0.016, Figure~3b for BH+0.019 vs. IR.

It is evident upon inspection that the distribution of reddening
residuals for the low reddening galaxies (Figures~1) are reasonably
Gaussian, consistent with the initial assumption that random errors 
dominate these data, and ex post facto justification of the
least-squares analysis performed here.  It is also evident that 
the reddening residuals for the high-reddened sample do not define 
a reasonable Gaussian sample (making their least-squares analysis a
bit suspect), suggesting that other issues might be at play when 
line-of-sight reddening is high.

The $1\sigma$ least-squares error for the relationship of Mg$_2$ on
(B-V)$_0$ is 0.027 mag for the low reddening galaxies (lines 2, 5, 8,
and 11 of Table~2)), as well as for the whole 378 galaxy sample (lines
1, 4, 7, and 10 of Table~2).  Consistency of errors determined for the
Mg$_2$-(B-V)$_0$ relation for all low reddening subsets of these data
further justifies the Gaussian assumption for the least-squares fits
to these data.

Galaxies with high Galactic reddening are generally situated at low
Galactic latitudes and/or in regions with irregular stellar
foregrounds. This observer can attest to the difficulties in obtaining
reliable (B-V) colors for such galaxies. Hence, a larger scatter in
the Mg$_2$-(B-V)$_0$ relation for high reddening galaxies was expected
(and is the reason for dividing the data set into low and high
reddening areas).

The $1\sigma$ agreement in reddening between the BH and IR methods 
for 378 gE galaxies is either 0.0144 mag (IR-0.016:BH) or 0.0158 mag
(BH+0.019:IR), for an average $1\sigma$ scatter between these
reddening predictions of 0.0151 mag.  The $1\sigma$ errors of
reddening predictions in low reddening directions are 0.006-0.009 mag
for the IR method, and 0.011 mag for the BH method, with BH and IR
predictions methods agreeing to an accuracy of 0.007 mag in these
directions. In the high reddening areas (but excluding the
``southern'' region with high gas-to-dust ratios), errors remain 
0.011 mag for the IR method but increase to 0.024-0.025 mag for the 
BH method. Line 2 of Table 2 also shows some evidence of skewness 
in the IR predictions for high reddening areas, yielding a mean of 
0.005 mag when the median is chosen to be zero; the BH values for 
high reddening areas show negligible skewness.  

Taking the 378 galaxy sample as a whole, the IR values yield more
accurate reddenings based on these elliptical galaxy data than do the
BH predictions, 0.007-0.009 mag to 0.013 mag.  So, while Schlegel
et al. were correct in stating that the IR method is more accurate
than the BH method, in reality it is a difference in reddening predictions
over most of the sky that is measured in millimags (i.e., a difference 
generally not detected by users of either method).

\section{ISM Issues and gE Stellar Populations}

\subsection{Polar reddening}

\citet{bh82} give a major review of reddening at the Galactic poles,
in which they determined the direction of lowest reddening (found at 
the North Galactic pole) from the average reddening of BAF stars
determine by both the uvby$\beta$ method and the average interstellar
polarization of those stars. In turn, this direction of lowest 
reddening (consistent with zero reddening as measured by these 
indicators) was used to determine the BH zero point.  In contrast, 
\citet{sfd98} determine the average reddening at the NGP by
directly measuring infrared emission, and assuming a one-to-one 
relation between IR dust emission (modulated by dust temperature) and 
reddening.  Hence, in both methods, the zero points assumed and the 
average value of reddening at the Galactic poles are linked.

Since \citet{bh82}, various authors have tried to show that the 
reddening at the NGP is more than obtained in that review,
Schlegel et al. included.  Schlegel et al. give two arguments 
in support of their finding non-zero reddening at the NGP: First, 
they cite the study of \citet{t90}, which claims that low
reddening at the NGP is an artifact due to stellar samples which 
select against more reddened stars (a type of selection effect 
previously pointed out by \citet{bh78a} in connection with galaxy 
selection issues; see \S 3.6). Second, Schlegel et al. suggest that 
dust ablation that could occur at the NGP would not eliminate 
``the small grains that dominate the UV and optical extinction.''

The Teerikorpi analysis leaves several issues unresolved, namely:

1. The fact that both BH and IR methods predict some measurable 
reddening (E(B-V)$\sim$ 0.01-0.03 mag) within 15$^\circ$ of the NGP, 
and that this reddening is likely at a distance of 150-250 pc from us 
\citep[cf.][]{bh78b} is ignored. No check was made if the reddening 
shown in Figure~2 of \citet{t90} is correlated in position on the sky 
or if directions with zero reddening are also correlated on the sky.

2. A claim is made that predicted E(b-y) is higher for more luminous,
bluer (and more distant) AF stars.  However, this effect is seen for
all of the distance bins in Figs. 4 and 5 of that paper, including
that for the nearest stars (for which no selection bias is claimed in
Teerikorpi's analysis).  This suggests more a calibration error in the
reddening predictions for the bluest stars in his sample, rather than
the kind of reddening-dependent selection effect proposed by
Teerikorpi.  This calibration error would also explain the higher
reddening that Teerikorpi obtains from the uvby$\beta$ data than was
obtained by \citet{bh82}.

3. It is stated that there is a general reddening of E(B-V) = 0.04 
mag over the whole NGP. Yet, the average E(B-V) determined by the 
IR method is 0.015 mag, averaged over 10$^\circ$ around the NGP
\citep{sfd98}, and that of BH is consistent with this picture,
but at a bit less reddening.

\citet[cf.][]{bt02} have also recently published new polarization 
measurements of stars at the NGP. Their measurements indicate 
large regions around the NGP where no polarization of starlight is
measured, consistent with the directions absent of stellar polarizations
found by \citet{ap75} and \citet{mark79} (the data used by BH in
1982 to help establish their zero point).  The new stellar
polarization measures of \citet{bt02} actually strengthen the 
result that the NGP contains regions with line-of-sight
 E(B-V)$<<0.01$ mag.

It is relevant here that \citet{bh78b} and \citet{bh82} point out that high
gas-to-dust ratios exist preferentially as halos around nearby OB
associations at low latitudes.  (Those high gas-to-dust ratios
{\it cannot be due} to either molecular clouds, or missing HI flux in
the HI maps used by Burstein \& Heiles (a suggestion made by
\citet{betal96}), as both effects would lower the gas-to-dust ratio,
not raise it.)

Not all the dust that produces emission produces reddening in the UV
and optical \citep[cf.][]{lg97}; a point also made by Schlegel et al.
However, the halos of the OB associations tell us that the outflows
from these associations can diminish the reddening from the dust while
not diminishing the HI gas.  It is known that our solar system is
passing through an HI superbubble of the type originally identified by
\citet{ch79} \citep[cf.  discussion in][]{sbpo85}.  Hence, our Sun is
now within a region that is analogous to the outflow halos of the OB
associations, so by analogy, high gas-to-dust ratios are likely
at high Galactic latitudes. \citet{gl95} suggest that high gas-to-dust
regions could be due to the outflows whittling down the ice mantles on
the dust such that more small grains are produced relative to larger
grains, reducing the UV and optical extinction.

In summary, the NGP has a patchy distribution of low amounts of
reddenings, including directions where E(B-V)$<<0.01$ mag with no
measurable polarization of the stars.  Yet, in all directions in the
NGP, there are quite measurable amounts of HI column density. The
conclusion arrived at by BH in 1982 still stands, namely that the 
NGP is a region where the gas-to-dust ratio is high.

\subsection{Gas-to-dust variations}

The dust-to-gas map (Figure~11) in \citet{sfd98} shows gas-to-dust
ratio variations of a factor of 3, both high and low.  This is 
consistent with what was found in \citet{bh78b}. Schlegel et al. 
also find there are significant gas-to-dust ratio variations in 
directions where they also find low reddenings (their Table~5).  
Among the 9 of the 18 directions with HI measures, the HI-to-dust 
emission ratio varies by nearly a factor of three. Hence, gas-to-dust 
variations along lines-of-sight in our Galaxy are real and pervasive.

Partly to study gas-to-dust variations, \citet{betal96} combined their
Leiden/Dwingeloo HI survey with COBE IR data, but restricted their
analysis to regions with HI column densities of less than $5 \times
10^{20}$ cm$^{-2}$ (224 BH units of HI column density). This limits
their investigation to northern Galactic latitudes $>
65^\circ$. \citet{betal96} find no variation in gas-to-dust ratio in
their analysis. Yet, it is precisely at high Galactic latitudes that
the BH method assumes a constant HI gas-to-dust ratio to predict
accurate reddenings. Hence, there is no disagreement between the BH
conclusions and those of Boulanger, et al. about the relative
constancy of gas-to-dust ratios at high northern Galactic
latitudes. And, conversely, the Boulanger et al.  analysis says
nothing about gas-to-dust ratio variations in the Galaxy at lower
Galactic latitudes

It is evident that something has to happen to the dust to make its
reddening characteristics get lower relative to the HI gas when it is
found in regions of outflow around OB associations. Given that the
most likely origin of Heiles-type HI superbubbles are interactions
with outflow winds of OB stars and supernovae, it is likely that the
dust emission seen at high galactic latitudes has the same
reddening-reduced characteristics as that around the OB associations.
This then would explain the observations of high gas-to-dust ratios
for high latitude dust.

\subsection{Zero Points of the Reddening Predictions vs. Minimum
Reddening Lines-of-Sight}

The BH method predicts large sections at high Galactic latitudes 
where reddening is predicted to be zero.  Given the estimated error 
in BH zero point (see above), this really means lines-of-sight 
where E(B-V)$<0.01$ mag. Eighteen (18) directions in Table~5 of 
Schlegel et al. yield line-of-sight E(B-V) estimates of 0.003 to 
0.015 mag; most have E(B-V)$<<0.01$ mag. Among the 402 gE there
are 62 with low IR-predicted E(B-V), including five with 
E(B-V)$\le 0.008$ mag.  

The fact that the zero point of the IR method is 0.02 mag higher than
that of the BH method does not prevent the IR method from predicting
reddenings with E(B-V)$\le 0.015$ mag in many directions on the sky.
In other words, the zero point difference between the IR and BH
methods is {\it not} due to the IR method predicting at least 0.02 mag
of E(B-V) reddening all over the sky.  Rather, this zero point
difference arises from the different assumptions made by both BH and
IR on how their reddening indicators correlate with reddening.  The
specific issue of the IR zero point is considered in detail below (\S
4). Here we note many papers have pointed out that if line-of-sight
reddening is always non-zero, a wide range of astrophysical issues are
affected. Not the least of these are the ages of Galactic globular
clusters \citep{vbs90}, as for every 0.01 mag that the E(B-V) values
are wrong, age errors of 1-2 Gyr are made.

\subsection{Significant Differences and Two Specific Directions}

In the difference map of BH and IR reddening predictions 
(Figure~12 in Schlegel et al.) are areas, mostly at low Galactic
latitudes, where the two sets of predictions vary significantly.
However, it is not sufficient to simply calculate the differences in
reddening predictions for the BH and IR methods, as illustrated by the
development of the HI-to-reddening relations in the BH papers.  One
needs to have independent reddening estimates to identify which, if
either method, is in error.  

Figure~5 plots the difference in BH and IR predictions versus the
residuals from the (B-V)$_0$-Mg$_2$ relation for the gE galaxies.
Lines are drawn at the $\pm 0.0453$ mag level (i.e., $3 \times 0.0151$
mag) to identify those galaxies for which the BH and IR reddening
predictions differ significantly.  Nine galaxies, mostly in
high-reddening directions, have predicted reddenings that differ by
more than 0.043 mag (see Table~3).  Interestingly, for all but one 
of these 9 galaxies, the IR reddening is significantly higher than 
the BH reddening.  Inspection of Table~3 shows that four galaxies 
have IR reddening predictions better than BH predictions, four galaxies
have BH predictions better than IR predictions, and for one galaxy 
both predicted residuals are within $3\sigma$, but of opposite sign.  

Two specific regions in the sky were initially singled-out for special
examination: the Perseus galaxies and the 11 galaxies in the
``southern region'' specified above. Figure~4 shows the predictions 
of the BH and IR methods differ systematically for the Perseus 
galaxies (cf. Table~3) which all lie in the same direction of the sky.  
Examination of BH and IR predictions for the (B-V)$_0$-Mg$_2$ 
relation for these eight galaxies reveals that the IR prediction has 
smaller errors for six of the eight galaxies. However, for three of 
the galaxies both IR and BH errors are large.  The Perseus galaxies 
are among the lowest-latitude, most heavily-reddened galaxies in the 
7 Samurai sample.  It is worth emphasizing again the observational 
difficulties of obtaining accurate B and V magnitudes for galaxies 
in highly-reddened, crowded stellar fields.

The 11 highly-reddened galaxies in the region $230^\circ > l >
310^\circ$, $-20^\circ < b < 20^\circ$ had their BH reddening
predictions halved for this paper. In contrast to the
``note-added-in-proof'' in the \citet{sfd98} paper, the adjusted BH
predictions are in generally good agreement with gE galaxy colors
(cf. Table~3). It is also noted that only 3 of these galaxies are in
the much smaller region $230^\circ > l > 240^\circ$, $-15^\circ < b <
-10^\circ$ pointed out in the \citet{sfd98} paper as having HI
observational problems. The differences in predicted reddenings
between BH and IR methods for these 11 gE galaxies will be discussed
in detail in \S 4.

\subsection{The stellar populations of gE galaxies}

(B-V)$_0$-Mg$_2$ relation errors include three sources: observational
errors for (B-V) and Mg$_2$; cosmic scatter among these stellar
populations; and cosmic scatter in the gradients of these stellar
populations {\it within} gE. At minimum, one has to assume an average
observational error in Mg$_2$ of 0.01 mag (cf. Davies et al. 1988),
and an observational error of 0.02 mag in (B-V) (as the original
estimate of 0.03 mag in Burstein et al. 1987 cannot be correct). One
is therefore left with a residual cosmic scatter of only 0.011
mag. This small cosmic scatter shows us that the stellar populations
of gE, at least as measured through the (B-V) color and the Mg$_2$
index, are remarkably uniform both in overall terms, as well as in
their internal gradients (see \citet{trager00} for reasons why this
might be the case).  Couple this with the strong relation of stellar
population with kinematics found in \citet{bbf93}, and one has some
very severe constraints on how the family of elliptical galaxies has
to be formed.

Of the 344/351 galaxies in low reddening directions, six galaxies have
good agreement in BH and IR reddening predictions, but have $\ge 0.07$
mag deviations in their (B-V)$_0$ colors, having them both too blue (2
galaxies) and too red (4 galaxies) for their Mg$_2$ measurements (see
Table~3). The two galaxies which are too blue for their Mg$_2$ values
have BH and IR reddening predictions that are too small to compensate 
for the blueness of their (B-V)$_0$ colors.  Five of these galaxies 
have close, early-type galaxy companions
\citep[cf.,][]{t88,smfetal89}, while the sixth galaxy is in the Coma
cluster (for which a close companion is hard to determine).  It would
be worthwhile to study the stellar populations in these 6 galaxies in
more depth.

\subsection{Why the IR method should be more accurate than the BH
Method}

In \citet{bh78a} was concerned with the result of \citet{as73,as75},
who obtained a csc-law slope that was smaller (0.03 mag) in E(B-V)
than that obtained by others at that time (0.05 mag). What
\citet{bh78a} found is the gE at low Galactic latitudes used by
Sandage to calibrae his csc-law (ironically, in the Perseus cluster)
lie preferentially in areas of low reddening compared to other
lines-of-sight in the same region.  In choosing galaxies (or stars)
for study, one chooses brighter galaxies (stars) which lie 
preferentially in directions of lower Galactic reddening.

Now couple this selection effect with the fact that the distribution
of dust is very patchy on all size scales on which it has been 
measured \citep[cf.][]{sfd98}, and the levels of patchiness increase
and the size scales of the patchiness decrease as reddening
increases.  As a result, the predictive power of any line-of-sight 
reddening prediction of finite pixel size will deteriorate in 
accuracy as the line-of-sight reddening increases, and/or with 
larger pixel size.  

On average, galaxies (or stars) chosen on the basis of their
brightness will be chosen in regions that have less reddening than the
average around them. This effect is minimal for low-reddening
directions, where variation in reddening is mostly comparable to or
larger than either the BH or the IR pixels, but becomes important for
highly-reddened regions, where patchy reddening exists on small
angular size-scales.  Hence, there will be a tendency for both the BH and
the IR methods to overestimate the reddenings of highly-reddened
galaxies (i.e., produce galaxies with colors too blue), with the onset
of these systematic effects occurring at lower reddenings for the 
larger BH prediction pixel than for the smaller IR pixel.  This is 
likely the primary reason why the IR predictions are more accurate 
than the BH predictions in highly-reddened directions, and slightly 
more accurate in low-reddening directions.

Inspection of Figures~2 shows that for E(B-V)(Ave) above 0.15 mag,
there is a tendency for the BH method to, indeed, overpredict the
reddenings of the galaxies (makes the galaxies too blue).  Inspection
of Figure~6 in \citet{sfd98} shows that for reddenings larger than
E(B-V) = 0.20 mag, the IR-predicted values also systematically
overestimate the observed reddenings. This is yet another reason for
one to be careful if one wishes to extend either of these reddening 
prediction methods to high-reddening directions.

Many of the 7 Samurai galaxies with high reddenings are south of $\rm
\delta = -23.5^\circ$, and only have HI predictions from the BH
method. The fact that there is no zero point shift, even for the 11
``southern'' gE whose HI-predicted BH reddenings have been halved (see
\S 4) indicates good consistency of HI measures, north and south.

\subsection{Completeness Issues}

BH: Areas in the south Galactic polar cap are missing BH predictions,
as well as some directions elsewhere in the sky.  The region
$230^\circ > l > 310^\circ$, $-20^\circ < b < 20^\circ$ needs special
attention (see above). The reddening for M31 is that of M32 in the BH
method, as the HI of M31 interferes with the measurement of HI in our
Galaxy.

IR: There might be an overestimate of its zero point and higher
scatter in their predictions related to high gas-to-dust ratios (\S
4).  No reddening prediction is made for M32, as the contribution of
background M31 flux is uncertain (the opposite problem for this pair
of galaxies than that faced by the BH method).  The $\rm 5\%$ of the
sky not covered by the IRAS survey will have lower spatial resolution
of the COBE/IR prediction than the rest of the sky.

\section{The IR:BH Zero Point Difference: A Matter of High Gas-to-Dust 
Ratios?}

Examination of the Schlegel et al. paper finds that the 
dust:HI emission ratio image (their Figure~11) looks very similar 
to the IR:BH difference image (their Figure~12).  Superimposing 
these two images confirms this impression: even very fine structure 
in the HI:dust emission map is the same as that in the IR:BH map. 
However, if we compare Figure~11 of Schlegel et al. to the
relevant figures in Heiles (1976) \citep[cf., discussion in][]{bh78b}, 
then a curious difference emerges. The regions that are low 
gas-to-dust ratio in the Heiles maps correspond to regions where 
there is active star formation, and the regions of high gas-to-dust 
ratio surround these regions. As discussed in the BH paper, this
makes sense if the high gas-to-dust ratio gas is due to the outflows
from the star-forming regions. In contrast, in Figure~11 of 
Schlegel et al. the opposite occurs, as areas of {\it higher} 
gas-to-dust ratio tend to appear on that map {\it inside} 
regions of {\it lower} gas-to-dust ratios.  This discrepancy could
be solved if Figure~11 of Schlegel et al. is really a map of
gas-to-dust ratios, rather than what they advertise it to be.
For now, however, this discrepancy stands.

There are two regions sampled by the gE galaxies where it is known
a priori that the HI/dust ratio is high: around the North
Galactic Pole and the 11 ``southern'' region $230^\circ > l > 310^\circ$,
$-20^\circ < b < 20^\circ$).  Borrowing on the correlation seen for
Figures~11 and 12 of Schelgel et al., their Figure~12 shows this
``southern'' region has the signature of patchy high gas-to-dust ratios. 
These 11 galaxies are listed in Table~3 (all 11 have an asterisk 
in their names).

Given that this is the only region for which there are both
highly-reddened galaxies, and high gas-to-dust ratios, it is of
interest to separately calculate the IR and BH reddening prediction
errors for these particular 11 galaxies (these were not included in
the error determination previously done). The value $\rm
\sigma(IR:BV)$ is 0.0337 for the IR reddening predictions for these 11
galaxies, significantly higher than 0.0212 found for $\rm
\sigma(BH:BV)$.  While 0.0337 for $\rm \sigma(IR:BV)$ is not out of
line for IR values for high-reddened galaxies (cf. Table~2), the value
of 0.0212 for $\rm \sigma(BH:BV)$ is significantly smaller that its
values for high-reddened galaxies. Moreover, further inspection of
Table~2 shows that the values of $\rm \sigma(IR:BV)$ are always
smaller than those of $\rm \sigma(BH:BV)$.  It is in this one region
of space, where we have galaxies with high reddening {\it and} high
gas-to-dust ratios, that the value of $\rm \sigma(IR:BV)$ is much
larger than that of $\rm \sigma(IR:BV)$.

Equally important, the IR method has has a zero point overprediction of
0.0165 mag in E(B-V) for these 11 galaxies, relative to the BH
predictions (even after the average zero point difference between the
two methods is eliminated). For these 11 galaxies the average 
zero point difference between IR and BH predictions is 0.033 mag in
E(B-V), not 0.016 mag as found for the 378 gE sample (from which these
11 galaxies were excluded).  

As reviewed in \S 3.1, there is good reason to think that the 
Galactic poles are dominated by regions of high gas-to-dust ratios, 
similar to that found around OB associations, as we may be passing 
through the remnants of an old, Heiles-like superbubble in the ISM.  
As Schlegel et al. restricted their determination of the IR:BH zero point 
difference to directions with $\rm \mid \! b \! \mid > 45^\circ$), 
it is precisely around the Galactic poles that their zero point 
difference of E(B-V) = 0.02 mag was calculated.  Moreover, half
of the 378 gE sample lie above Galactic latitudes of 45$^\circ$
(which is less than 1/2 of the sky).

The current evidence indicates that in regions of higher reddening,
the IR minus BH zero point difference doubles to 0.033 mag. 
{\it Hence, if the IR method overpredicts E(B-V) in directions of 
high gas-to-dust ratios, with the size of the overprediction 
increasing as the degree of reddening increases, this would fully 
explain the zero point difference seen between the two methods.}

Appealing to Table 5 of Schlegel et al. for further information is of
no avail, as IR (and BH) predicted E(B-V) in the 9 directions in that
table having HI measures are $<0.015$ mag.  The mutual agreement
between the BH and IR methods, 0.007 mag in regions of low reddening,
is comparable to this range of E(B-V), making it fruitless to try to
use these 9 directions to see if a relation exists between gas-to-dust
ratio and the IR vs. BH zero point. One needs more highly-reddened,
high gas-to-dust directions to make this test, the only such region in
the present dataset is the southern hemisphere region discussed here.

The principal difference between BH and IR methods is that the BH
method explicitly accounts for dust-to-gas ratio differences at low
Galactic latitudes, while the IR method does not.  As discussed in the
previous section, there is reason to believe that variations in
gas-to-dust ratio along different lines-of-sight in the Galaxy (which
the IR method {\it does} confirm exist) are due to the modification of
the dust grains in outflows from OB stars and supernova
\citep[cf.][]{gl95}.

Thus, there are two lines of evidence that strongly suggest the IR
method overestimates reddenings in directions with high gas-to-dust 
ratios: First, the IR method has a 0.016 mag in E(B-V) higher 
zero point than the BH method as primarily determined in regions 
of low reddening, with half the sample near the Galactic poles. 
Second, the IR predictions overestimate reddenings at twice 
this level relative to those from the BH method for highly-reddened 
galaxies in the one low latitude, high gas-to-dust region that is 
sampled by the 7 Samurai data set.  A third, related issue is
the close correlation of fine-leveled structure in the IR:HI and 
IR:BH maps in \citet{sfd98}. 

It is worth reflecting on the fact that to further test the IR
predictions in regions of high gas-to-dust ratio and of moderately
high reddenings, one is restricted to directions at relatively low
latitudes. Yet, very few galaxy surveys include galaxies at low
latitudes in their samples (unless one is probing the Zone of
Avoidance itself).  That there are any gE galaxies at low Galactic
latitudes in the 7 Samurai sample is an accident of galaxy
selection. Rather, most investigations of galaxies avoid
highly-reddened regions more because they lie in crowded stellar
regions than they lie in highly-reddened directions (cf. the
relatively high Galactic latitude limits of the Sloan Digital Sky
Survey).  Also due to the accident of their selection (from the ESO
survey plates), most of the highly-reddened galaxies in the 7 Samurai
survey lie below declination -23.5$^\circ$, the nominal northern
galaxy count survey limit.  Unless a concerted effort is made to study
gE galaxies at low northern hemisphere latitudes (with their
concomitant observational problems) in directions where there is known
high gas-to-dust ratios, further probing of the issue of high
gas-to-dust dependency of the IR method (and the associated diminished
reddening of dust relative to overall dust emission in those
directions) will be difficult.

\section{Summary}

Both \citet{bh78b} (the BH method) and \citet{sfd98} (the IR
method) permit one to independently predict line-of-sight Galactic
reddening.  Both methods can be differentially calibrated using
independent reddening estimates for objects situated beyond the
Galactic reddening layer.  The currently best data for this
exercise are the (B-V)$_0$ colors and Mg$_2$ line strengths for
giant elliptical (gE) galaxies, as it is found empirically 
\citep{dbetal88} that their stellar populations are remarkably 
uniform for these these parameters.  This paper presents a 
revised 7 Samurai database from which reddening predictions can be 
made more accurately for 402 gE galaxies.  Of these 402 gE galaxies, 
12 lack direct BH predictions, one lacks direct IR prediction, and 
11 galaxies lie in the southern region identified in \citet{dbetal87} 
as needing an adjustment to the BH predictions ($230^\circ > l > 
310^\circ$, $-20^\circ < b < 20^\circ$), leaving 378 galaxies for an 
objective, least-squares analysis.

Both BH and IR methods must assume zero points for their predictions.
The BH method uses independent estimates of E(B-V) at the NGP and
takes its zero point from the directions with the least amount of
reddening and minimum amount of average stellar polarization (E(B-V)
$\sim$ 0.00). The IR method assumes that there is a one-to-one
relationship between dust emission/temperature and reddening. Because
the IR method sees dust emission at the poles, the BH and IR methods 
differ in zero point by $\sim 0.02$ mag. The present analysis is 
consistent with this finding: IR zero point higher by 0.016 mag when 
the BH zero point is applied to the IR data, and higher by 0.019 mag 
when the IR zero point is applied to the BH data.

Two slopes are used for the (B-V)$_0$-Mg$_2$ relation, owing to 
differences in how this relation can be defined. The sample is 
further divided into low/high reddening directions, as well as 
taking the whole dataset together (taking the low/high reddening 
cut at E(B-V) = 0.100).  For each of these twelve data combinations, 
a system of four coupled equations in four unknowns are formed that 
can be solved in a least-squares manner for the intrinsic $1\sigma$ 
errors for four quantities of interest: BH-predicted reddenings; 
IR-predicted reddenings; the agreement between the BH and IR 
reddening predictions, and the (B-V)$_0$-Mg$_2$ relation itself.

In directions with E(B-V)$<$0.10 mag, IR prediction errors are
0.002-0.005 mag in E(B-V) more accurate than those of the
BH method, yielding an average $1\sigma$ prediction accuracy of
0.007 mag.  The IR method is more accurate (0.011 mag) than the BH
method in most highly-reddened directions (0.024-0.025 mag).  Owing
to a selection effect first detailed by \citep{bh78a}, the higher 
accuracy of the IR method is likely related to its smaller pixel 
size on the sky (6$'$) compared to that of the BH method ($18'
\times 36'$).  Observational errors in the (B-V)$_0$-Mg$_2$ relation
are 0.027 mag in low reddening directions and higher in high-reddening
directions, owing to observational issues in accurately measuring 
galaxy colors in crowded stellar regions. Only 6 galaxies, all 
likely companions of larger galaxies, are found in low-reddening 
directions for which their (B-V) colors are significantly too blue or
too red for their Mg$_2$ line strengths. These galaxies are worth 
further investigations.

Discussion of the reddening at the North Galactic Pole is updated 
from that given in \citet{bh78b}, but the same conclusion is 
drawn from newer data as was drawn then: there are directions 
around the NGP that have E(B-V)$<<0.01$ mag; i.e, essentially not 
measurable. Furthermore, a plausible case can be made that 
the Solar System is presently moving through an old Heiles-type 
superbubble \citep{ch79}, in which the dust has been affected by 
the same mechanism (a whittling-down of its ice mantles?; cf. 
\citet{gl95}) that produces high gas-to-dust ratios around 
OB associations (which are seen in both the BH and IR datasets).

IR and BH methods both find large variations in the HI-to-reddening 
ratio in the interstellar medium in our region of the Galaxy. High
ratios of gas-to-dust are found specifically around nearby OB 
associations and at the Galactic poles.  Both methods predict
many lines-of-sight where E(B-V)$<0.015$ mag, including five gE
for which the IR method predicts E(B-V)$<0.008$ mag.

There are two lines of evidence that the IR method overpredicts
reddenings by E(B-V)$\sim$ 0.02 mag in directions where there are high
gas-to-dust ratios: the much larger errors and larger E(B-V) zero
point (0.033 mag) of the IR method relative to the BH method for 11
highly-reddened gE galaxies in directions known to have high
gas-to-dust ratios; and the existence of the E(B-V)$\sim$0.02 mag zero
point difference determined primarily at high Galactic latitudes,
which are also directions of high gas-to-dust ratios.  Another set of
data which could bear on this issue, that of the IR:HI map published
in Schlegel et al., seems to be at odds with a similar HI:dust map
published in \citet{h76}, and used by BH to correct for gas-to-dust
ratio differences in their reddening map.  While structure in the
IR:HI difference map of Schelgel et al.  corresponds well to similar
structure in the IR:BH difference map they publish, the issue as to
what the IR:HI map means needs to be resolved before this comparison
can be used to help understand zero point difference between the IR
and BH methods.

Given the good differential agreement between the BH and IR
predictions for most directions where E(B-V)$<$0.100 mag, it is 
wise to compare the reddening estimates one gets from both
methods to discover the systematic errors in both methods. In so
doing, one can find out further information about what portions of the
dust are doing the reddening, as this information is currently
unknown. For example, the recent study of \citet{mcetal01} found that
taking an average of the two methods yielded a more accurate result
than from either method alone.

As long as there was one accurate method for predicting reddening
along lines-of-sight (BH) in the Galaxy, the issue of the assuming the
zero point of that method was apparently well-hidden from sight.  The
fact that there are now two accurate methods for predicting
line-of-sight reddening in the Galaxy that have assumed different zero
points gives us both the chance to learn more about the interstellar
medium, but also places the onus of choosing a common zero point upon
the user of either or both methods.  Inspection of the reddenings from
the two methods currently listed on the NED website brings this issue
to the forefront.  Having two accurate methods of predicting
line-of-sight reddenings in our Galaxy gives one the ability to
compare them in all directions that are used.  Given that systematic
differences do exist in these predictions, it is further recommended
that the values of the two predictions be compared for consistency
before averaging.

In closing, it is hoped that the Sloan Digital Sky Survey (SDSS) will
provide a new treasure-trove of E galaxy colors and spectral line
strengths which can be used to differentially test the BH and IR
reddening predictions in many different directions. Unfortunately,
even the SDSS will not probe into highly-reddened, high gas-to-dust 
regions, leaving those kinds of investigation for the future.  It is 
also hoped that Schlegel et al. will post files on their website that 
tabulate both the gas-to-dust ratios that their method calculates and
the IR to BH differences with the zero point difference between these
data removed.  Such tables would facilitate further studies of the
the ISM, as well as differences in reddening predictions between 
the BH and IR methods.

\acknowledgements

DB would like to thank the NED team, specifically Harold Corwin,
Marion Schmitz and Anne Kelly for providing their values for the
Schlegel et al. reddenings.  An anonymous referee is thanked for
helpful comments.  DB also thanks his colleagues in the 7 Samurai
collaboration for helping to develop one of the most accurate set of
data on elliptical galaxies that currently exists.

\clearpage

\begin{deluxetable}{lrrrrrrrrrrrrr}
\tabletypesize{\tiny}
\tablecaption{The Reddening Data}
\tablehead{
\colhead{Galaxy} & \colhead{Dec} & \colhead{l} &
\colhead{b} & \colhead{(B-V)} & \colhead{Mg$_2$} &
\colhead{BH} & \colhead{BH+} & \colhead{IR} &
\colhead{IR-} & \colhead{Dif1} & \colhead{Dif2} &
\colhead{Ave(BH+,IR)} & \colhead{Ave(BH,IR-)} }
\startdata
N~2012 & -79.89 & 292.02 & -30.54 & 1.10 & 0.287 & 0.111 & 0.130 & 
0.147 & 0.131 & -0.017 & -0.020 & 0.139 & 0.121 \\
N~6876 & -71.02 & 324.13 & -32.60 & 1.03 & 0.304 & 0.040 & 0.059 &
0.045 & 0.029 & 0.014 & 0.011 & 0.052 & 0.035 \\
N~2434 & -69.17 & 281.00 & -21.54 & 1.09 & 0.260 & 0.177 & 0.196 &
0.248 & 0.232 &-0.052 &-0.055 & 0.222 & 0.205 \\
N~3136 & -67.13 & 287.99 &  -9.45 & 1.03 & 0.280 &$\ldots$ &$\ldots$ &
0.238 & 0.222 &$\ldots$&$\ldots$&$\ldots$&$\ldots$ \\
N~3136B& -66.73 & 288.18 &  -8.81 & 1.12 & 0.272 &$\ldots$&$\ldots$&
0.198 & 0.182 &$\ldots$&$\ldots$&$\ldots$&$\ldots$ \\
N~7192 & -64.56 & 326.52 & -44.54 & 0.95 & 0.270 &-0.003 & 0.016 &
0.034 & 0.018 &-0.018 &-0.018 & 0.025 & 0.009 \\
N~2305 & -64.21 & 274.43 & -24.55 & 1.00 & 0.304 & 0.071 & 0.090 &
0.076 & 0.060 & 0.014 & 0.011 & 0.083 & 0.066 \\
N~6483 & -63.67 & 330.07 & -18.66 & 0.98 & 0.280 & 0.077 & 0.096 &
0.059 & 0.043 & 0.037 & 0.034 & 0.078 & 0.060 \\
N~2887 & -63.60 & 282.27 &  -9.64 & 1.12 & 0.260 &$\ldots$&$\ldots$&
0.225 & 0.209 &$\ldots$&$\ldots$&$\ldots$&$\ldots$ \\
N~6721 & -57.85 & 338.57 & -23.90 & 1.02 & 0.338 & 0.064 & 0.083 &
0.060 & 0.044 & 0.023 & 0.020 & 0.072 & 0.054 
 \enddata
\tablecomments{Column 1 is the name of
the galaxy as given in \citet{smfetal89}; Column 2 is its 1950
declination (needed to separate the BH subsets into north and south
subsets), by which this table is ordered; Columns 3 and 4 are its
Galactic longitude and latitude; Column 5 is its {\it observed} (B-V)
color, not corrected for reddening; Column 6 is its revised Mg$_2$
line-strength, described above; Column 7 is its predicted BH reddening
(negative values explicitly given); Column 8 is its predicted BH
reddening with 0.019 mag added to bring it into zero point agreement
with the IR method; Column 9 is its IR-predicted reddening as given by
Schlegel et al.; Column 10 is its IR-predicted reddening with 0.016
mag subtracted, making it consistent in zero point with the BH
prediction; Columns 11\&12 are its differences in E(B-V), BH+0.019 mag
minus IR and BH minus IR-0.016 mag, respectively (see definitions of
these terms below); Column 13 averages the BH+0.019 mag and the IR mag
predictions for this galaxy; Column 14 averages its BH and its
IR-minus-0.016 mag predictions. All predicted negative reddening
values, IR or BH, are set to 0.000 for reddening calculations. No
values are given in their respective columns when there is either no
BH or no IR prediction (these missing values are set to -0.999 in the
ASCII version of this table).\\
13 gE galaxies do not have direct BH or IR reddenings: six gE
galaxies have $\rm b < \mid \! 10^\circ \! \mid$, where BH predictions
do not exist (IC 2311, NGC 2380, NGC 2663, NGC 2887, NGC 3156, and NGC
3156B). Six gE galaxies are in various gaps in the BH reddening map where
HI data are absent, hence no direct BH predictions (NGC 641, NGC 822,
IC 1459, IC 1625, IC 5328, and ESO 409-G12).  NGC~221 (M32) has no
direct IR prediction.  A further 11 gE galaxies in the region 
$\rm 230^\circ < l < 310^\circ$, $\rm -20^\circ < b < 20^\circ$ were 
found in \citet{dlbetal88} and \citet{dbetal87} to have HI-only BH reddening
predictions that are a factor of two too high (NGC 2292, NGC 2293, NGC
2325, NGC 2888, NGC 2904, NGC 3087, NGC 3250, NGC 3557, NGC 4976, ESO
208-G21, and ESO 264-G31). (These galaxies are noted in this table by an
asterisk in their names.) Note that an additional 11 galaxies within 
a smaller region defined by $265^\circ < l < 310^\circ, 17^\circ < b < 
19.5^\circ$ (NGC 3108, NGC 3258, NGC 3257, NGC 3260, E376-G007, 
E318-G021, NGC 4946, NGC 5011, NGC 5090 and E270-G014) did {\it not} 
need to have their reddenings so-corrected in the 7 Samurai database.}

\end{deluxetable}

\clearpage

\begin{deluxetable}{lrrrrrrrrr}
\tabletypesize{\scriptsize}
\tablecaption{Least-Squares-Determined Observational and Intrinsic
  Errors. \label{tab2}}
\tablehead{ 
\colhead{Data Set} & \colhead{N} & 
\colhead{BH:BV} & \colhead{IR:BV} & \colhead{Ave:BV} 
& \colhead{BH:IR} 
& \colhead{$\rm \sigma(BH)$} & \colhead{$\rm \sigma(IR)$} 
& \colhead{$\rm \sigma(Ave)$} & \colhead{$\rm \sigma(BV+Mg_2)$} } 
\startdata
All:1.12:BH+& 378  & 306 & 293 & 288 & 158 & 128 &
93 & 79 & 278  \\
&& 22 & 31 & 21 & 16 &&&& \\
Low:1.12:BH+& 344  & 295 & 288 & 282 & 144 & 112 &
91 & 71 & 273  \\
&& 24 & 28 & 21 & 12 &&&& \\
High:1.12:BH+& 34  & 399 & 340 & 347 & 263 & 237 &
114 & 132 & 321  \\
&& 7 & 52 & 25 & 52 &&&& \\
&&&&&&&&& \\
All:1.12:IR-& 378 & 309 & 292 & 292 & 144 & 125 &
72 & 72 & 283  \\
&& 16 & 22 & 14 & -6 &&&& \\
Low:1.12:IR-& 351 & 298 & 286 & 285 & 129 & 110 &
69 & 65 & 277  \\
&& 17 & 19 & 13 & -10 &&&& \\
High:1.12:IR-& 27 & 427 & 361 & 372 & 266 & 248 &
98 & 133 & 348  \\
&& 3 & 59 & 26 & 43 &&&& \\
&&&&&&&&& \\
All:0.942:BH+& 378 & 297 & 282 & 279 & 158 & 130 &
91 & 79 & 267  \\
&& 0 & 18 & -1 & 16 &&&& \\
Low:0.942:BH+& 344 & 287 & 277 & 273 & 144 & 114 &
87 & 72 & 263  \\
&& 1 & 15 & -2 & 12 &&&& \\
High:0.942:BH+& 34 & 386 & 325 & 332 & 263 & 237 &
114 & 132 & 304  \\
&& -10 & 44 & 6 & 52 &&&& \\
&&&&&&&&& \\
All:0.942:IR-& 378 & 301 & 281 & 282 & 144 & 127 &
68 & 72 & 273  \\
&& -6 & 16 & 10 & -10 &&&& \\
Low:0.942:IR-& 351 & 291 & 275 & 276 & 129 & 113 &
63 & 65 & 268  \\
&& -6 & 16 & 10 & -10 &&&& \\
High:0.942:IR-& 27 & 408 & 345 & 354 & 266 & 244 &
107 & 133 & 327 \\
&& -15 & 60 & 28 & 43 &&&& 
\enddata
\tablecomments{Column 1 gives the type of sample used
(BH+ or IR- refers to the addition/subtraction of zero points for that
data set, 1.12/0.942 refers to the slope of the (B-V)$_0$-Mg$_2$
relation used, all/low/high refers to the subset of data used).
Column 2 gives the number of galaxies in this sample.  Columns 3-6
give the $1\sigma$ least-squares errors in E(B-V) found for the
observable errors (BH:BV, IR:BV, Ave:BV, and BH:IR, respectively), in
units of 0.0001 mag.  Columns 7-10 give the $1\sigma$ least-squares
errors derived from the least-squares solution of the above equations
for the intrinsic errors $\rm \sigma(BH), \sigma(IR), \sigma(Ave)$ and
$\rm \sigma(BV+Mg_2)$, in units of 0.0001 mag. The second line in
this table gives the zero point offsets calculated for each of the
least-squares fits under which they are placed.}

\end{deluxetable}

\clearpage

\begin{deluxetable}{lrrrrrrrrrr}
\tabletypesize{\tiny}
\tablecaption{Galaxies Separately Disucssed}
\tablehead{
\colhead{Galaxy} & \colhead{l} & \colhead{b} &
\colhead{E(B-V)$_{\rm BH}$} & \colhead{E(B-V)$_{\rm IR-}$} &
\colhead{E(B-V)$_{\rm BH+}$} & \colhead{E(B-V)$_{\rm IR}$} &
\colhead{Diff$_{\rm BH:IR}$} & \colhead{Diff$_{\rm BH}$} &
\colhead{Diff$_{\rm IR}$} & \colhead{Diff$_{\rm gal}$}}
\startdata
\multicolumn{11}{c}{Four galaxies for which IR reddenings more
accurate than BH reddenings} \\
&&&&&&&&&& \\
NGC 1060 & 148.77 &-24.79 &  0.117  &  0.193  & 0.136  &   0.209  &
-0.075 &  0.078 &  0.003 &  0.040 \\
NGC 1293 & 150.94 &-13.16 &  0.208  &  0.151  & 0.227  &   0.167  &
0.059 & -0.115 & -0.057 & -0.087 \\
NGC 1713 & 198.62 &-24.50 &  0.062  &  0.108  & 0.081  &   0.124  &
-0.045 &  0.061 &  0.017 &  0.039 \\
NGC 5017 & 310.32 & 45.80 &  0.024  &  0.068  & 0.043  &   0.084  &
0.043 &  0.053 &  0.010 &  0.031 \\
&&&&&&&&&& \\
\multicolumn{11}{c}{Four galaxies for which BH reddenings more
accurate than IR reddenings} \\
&&&&&&&&&& \\
NGC 821  & 151.55 &-47.56 &  0.040  &  0.094  & 0.059  &   0.110  &
-0.053 & -0.020 & -0.071 & -0.046 \\
E208*G021& 262.69 &-14.30 &  0.075  &  0.157  & 0.094  &   0.173  &
-0.081 &  0.015 & -0.066 & -0.027 \\ 
NGC 2434 & 281.00 &-21.54 &  0.177  &  0.232  & 0.196  &   0.248  &
-0.054 &  0.001 & -0.053 & -0.027 \\
NGC*4976 & 305.80 & 13.27 &  0.112  &  0.167  & 0.131  &   0.183  &
-0.054 & -0.027 & -0.080 & -0.054 \\
&&&&&&&&&& \\
\multicolumn{11}{c}{One galaxy for which BH and IR reddenings differ
  but both accurate} \\
&&&&&&&&&& \\
NGC*2888 & 257.15 & 16.08 &  0.057  &  0.108  & 0.076  &   0.124  &
-0.050 &  0.014 & -0.037 & -0.012 \\
&&&&&&&&&& \\
\multicolumn{11}{c}{Six galaxies for which BH and IR reddenings agree, but
  galaxy (B-V)$\rm _0$ deviates} \\
&&&&&&&&&& \\
NGC  83  & 113.86 &-39.91 &  0.032  &  0.054  & 0.051  &   0.070  &
-0.021 &  0.091 &  0.070 &  0.080 \\
NGC 708  & 136.58 &-25.08 &  0.060  &  0.072  & 0.079  &   0.088  &
-0.011 &  0.105 &  0.094 &  0.099 \\
NGC 1549 & 265.41 &-43.80 &  0.000  &  0.000  & 0.000  &   0.013  &
-0.007 & -0.069 & -0.076 & -0.073 \\
NGC 3610 & 143.54 & 54.46 &  0.000  &  0.000  & 0.013  &   0.010  &
-0.002 & -0.089 & -0.086 & -0.088 \\
NGC 4872 &  57.85 & 88.02 &  0.012  &  0.000  & 0.031  &   0.009  &
 0.017 &  0.082 &  0.098 &  0.090 \\
NGC 7617 &  87.64 &-48.35 &  0.040  &  0.069  & 0.059  &   0.085  &
-0.028 &  0.141 &  0.114 &  0.128 \\
&&&&&&&&&& \\
\multicolumn{11}{c}{The Seven Other Perseus Galaxies (in addition to
 NGC 1293)} \\
&&&&&&&&&& \\
IC 310    & 150.18 & -13.73 &   0.149 &  0.144 &  0.168 &  0.160 &  
 0.007 &  0.054 &  0.060 &  0.056 \\
NGC 1272  & 150.51 & -13.33 &   0.162 &  0.146 &  0.181 &  0.162 & 
 0.018 & -0.031 & -0.012 & -0.022 \\
NGC 1273  & 150.50 & -13.28 &   0.174 &  0.147 &  0.193 &  0.163 & 
 0.029 & -0.033 & -0.004 & -0.019 \\
NGC 1278  & 150.56 & -13.21 &   0.174 &  0.148 &  0.193 &  0.164 & 
 0.028 & -0.021 &  0.007 & -0.007 \\
NGC 1282  & 150.72 & -13.34 &   0.171 &  0.151 &  0.190 &  0.167 & 
 0.022 & -0.058 & -0.036 & -0.047 \\
NGC 1283  & 150.71 & -13.31 &   0.171 &  0.150 &  0.190 &  0.166 & 
 0.023 & -0.023 &  0.001 & -0.012 \\
CR 32     & 150.51 & -13.22 &   0.174 &  0.149 &  0.193 &  0.165 &  
 0.027 &  0.017 &  0.044 &  0.030 \\
&&&&&&&&&& \\
\multicolumn{11}{c}{The Additional Eight High Gas-to-Dust Galaxies in the 
$230^\circ > l > 310^\circ$, $-20^\circ < b < 20^\circ$ Direction} \\
&&&&&&&&&& \\
N*2292   & 236.73 &-12.61 & 0.141 &  0.103 &  0.160 &  0.119 &  
0.040 &  -0.018 &  0.022 &  0.002 \\
N*2293   & 236.74 &-12.60 & 0.141 &  0.104 &  0.160 &  0.120 &  
0.039 &  -0.031 &  0.008 & -0.011 \\ 
N*2325   & 239.96 &-10.41 & 0.140 &  0.101 &  0.159 &  0.117 &  
0.041 &  -0.016 &  0.025 &  0.005 \\
N*2904   & 259.52 & 15.06 & 0.083 &  0.110 &  0.102 &  0.126 & 
-0.026 &   0.006 & -0.021 & -0.008 \\
N*3087   & 266.89 & 16.33 & 0.072 &  0.090 &  0.091 &  0.106 & 
-0.017 &   0.011 & -0.007 &  0.002 \\
N*3250   & 274.97 & 14.93 & 0.066 &  0.087 &  0.085 &  0.103 & 
-0.020 &   0.011 & -0.009 &  0.001 \\
N*3557   & 281.58 & 21.09 & 0.069 &  0.083 &  0.088 &  0.099 & 
-0.013 &   0.033 &  0.020 &  0.026 \\
E264*G031& 280.53 & 10.91 & 0.100 &  0.139 &  0.119 &  0.155 & 
-0.038 &   0.043 &  0.005 &  0.024
 \enddata

\tablecomments{Column 1 gives the galaxy name; Columns 2 and 3
give the Galactic longitude and latitude of this galaxy; Columns 4-7
give, respectively, the BH reddening, the IR-0.016 reddening, the
BH+0.019 reddening and the IR reddening predicted for this galaxy;
Column 8 gives the difference, BH-IR, taking the average of the
differences BH minus IR-0.016 and BH+0.019 minus IR; Columns 9-11 
give, respectively, the difference in (B-V)$_0$ for this galaxy from 
the (B-V)$_0$-Mg$_2$ relationship using: a) the average of the 
two BH predictions (BH, BH+0.019); b) the average of the two 
IR predictions (IR, IR-0.016), and c) the average difference overall
by combining all four predictions.}

\end{deluxetable}

\clearpage

\begin{figure}
\epsscale{0.7}
\plotone{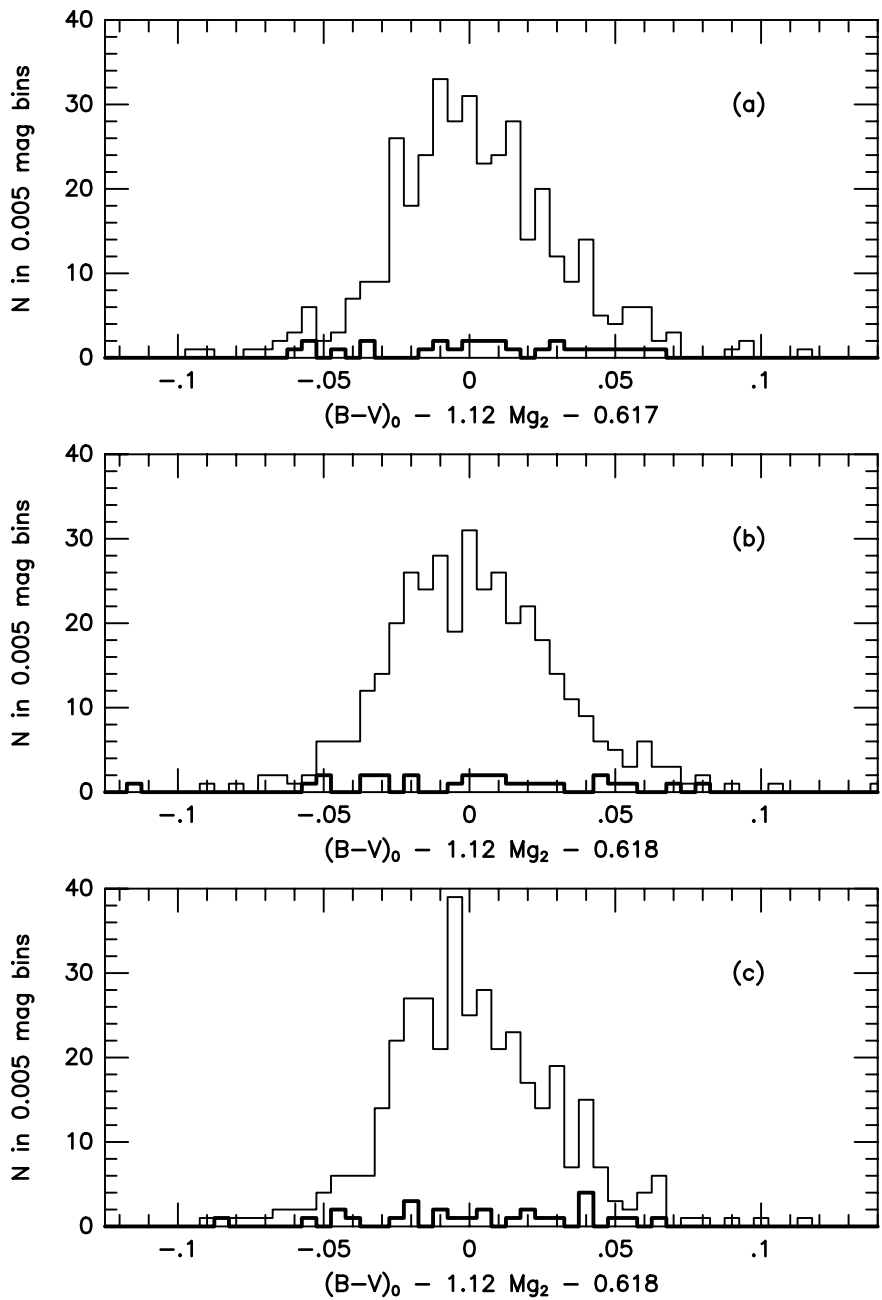}
\figcaption{The histograms of residuals from the (B-V)$_0$-Mg$_2$
relation for the 378 galaxy sample using the 1.12 slope for this
relation. The histograms for the highly-reddened galaxies are given in
darker lines: (a) the IR-0.016 predictions; (b) for the straight BH
predictions; (c) for the average of IR-0.016 and BH predictions.
\label{fig1a}}
\end{figure}

\begin{figure}
\epsscale{0.7}
\setcounter{figure}{0}
\plotone{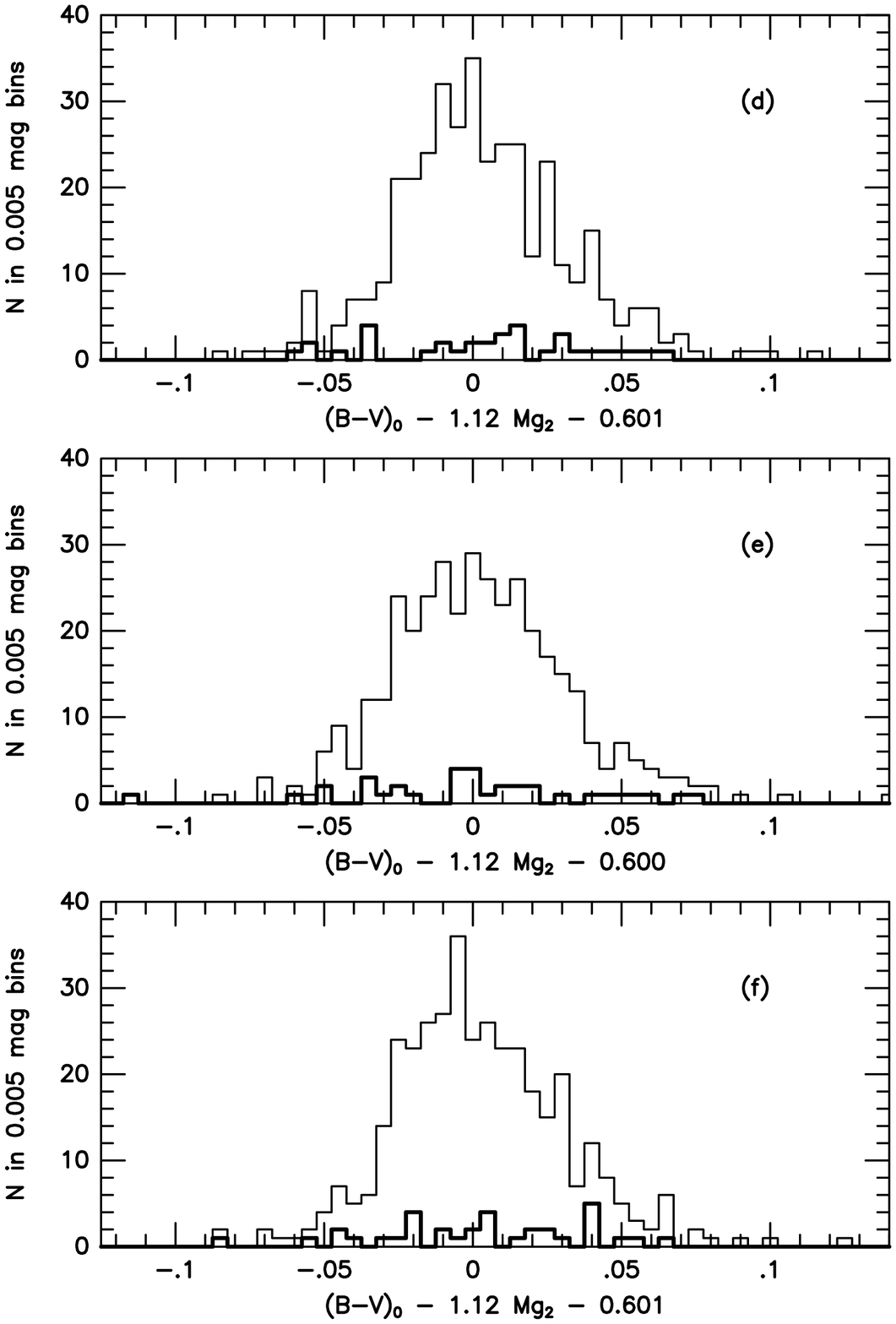}
\figcaption{(continued): (d) for the straight IR predictions; 
(e) for the BH+0.019 predictions; and (e) the average of IR and 
BH+0.019 predictions. \label{fig1d}}
\end{figure}

\begin{figure}
\epsscale{0.7}
\plotone{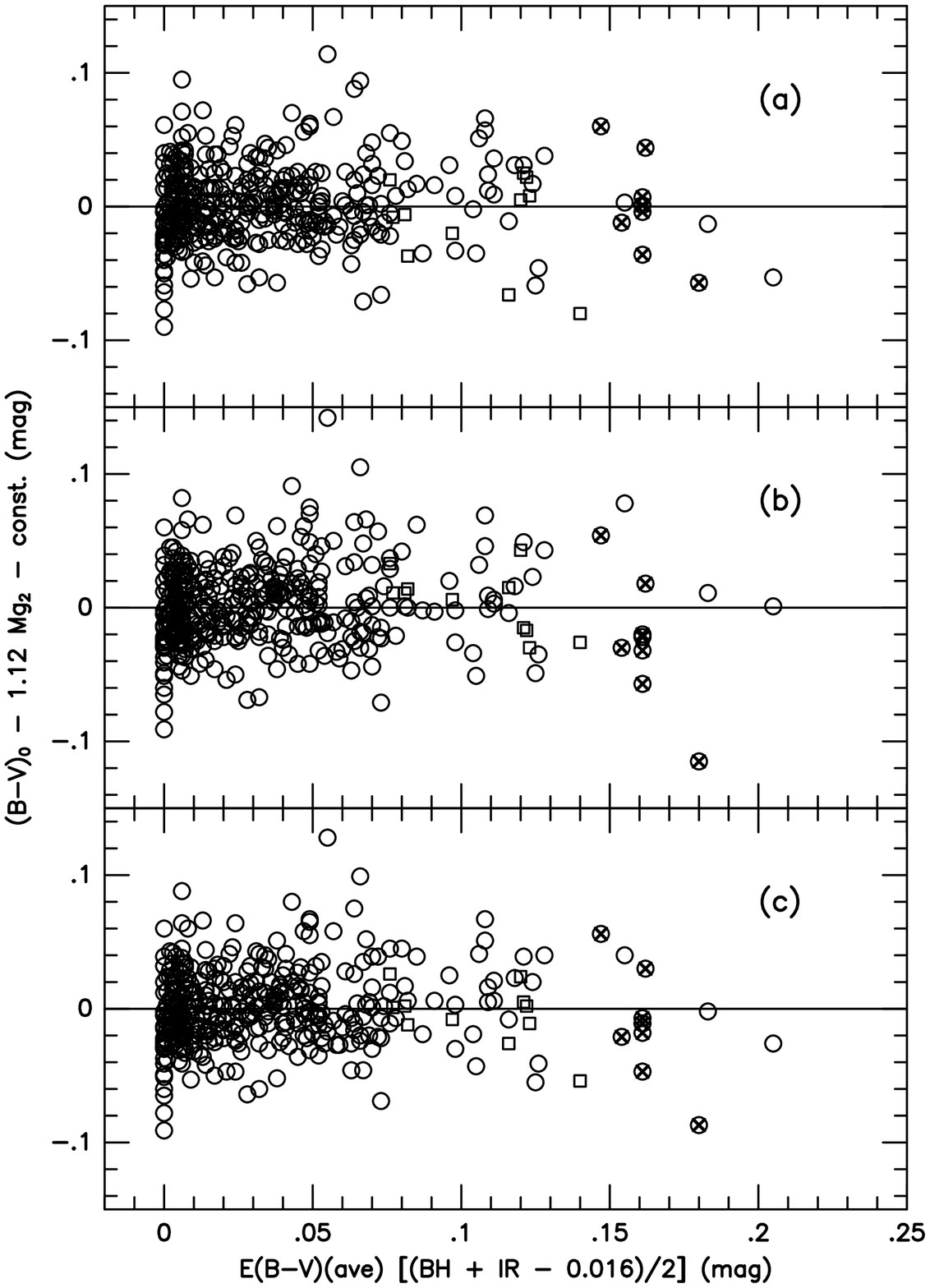}
\figcaption{The residuals from the (B-V)$_0$-Mg$_2$ relation plotted
versus the Ave value of E(B-V) for each galaxy: (a) the IR-0.016
predictions; (b) for the BH predictions; (c) for the average of
IR-0.016 and BH predictions.  Data for galaxies in the Perseus cluster
are plotted as circled X's; data for the 11 highly-reddened galaxies in the
region $230^\circ > l > 310^\circ$, $-20^\circ < b < 20^\circ$ are
added to this figure and are plotted as open boxes in each graph. 
\label{fig2a}}
\end{figure}

\begin{figure}
\setcounter{figure}{1}
\epsscale{0.7}
\plotone{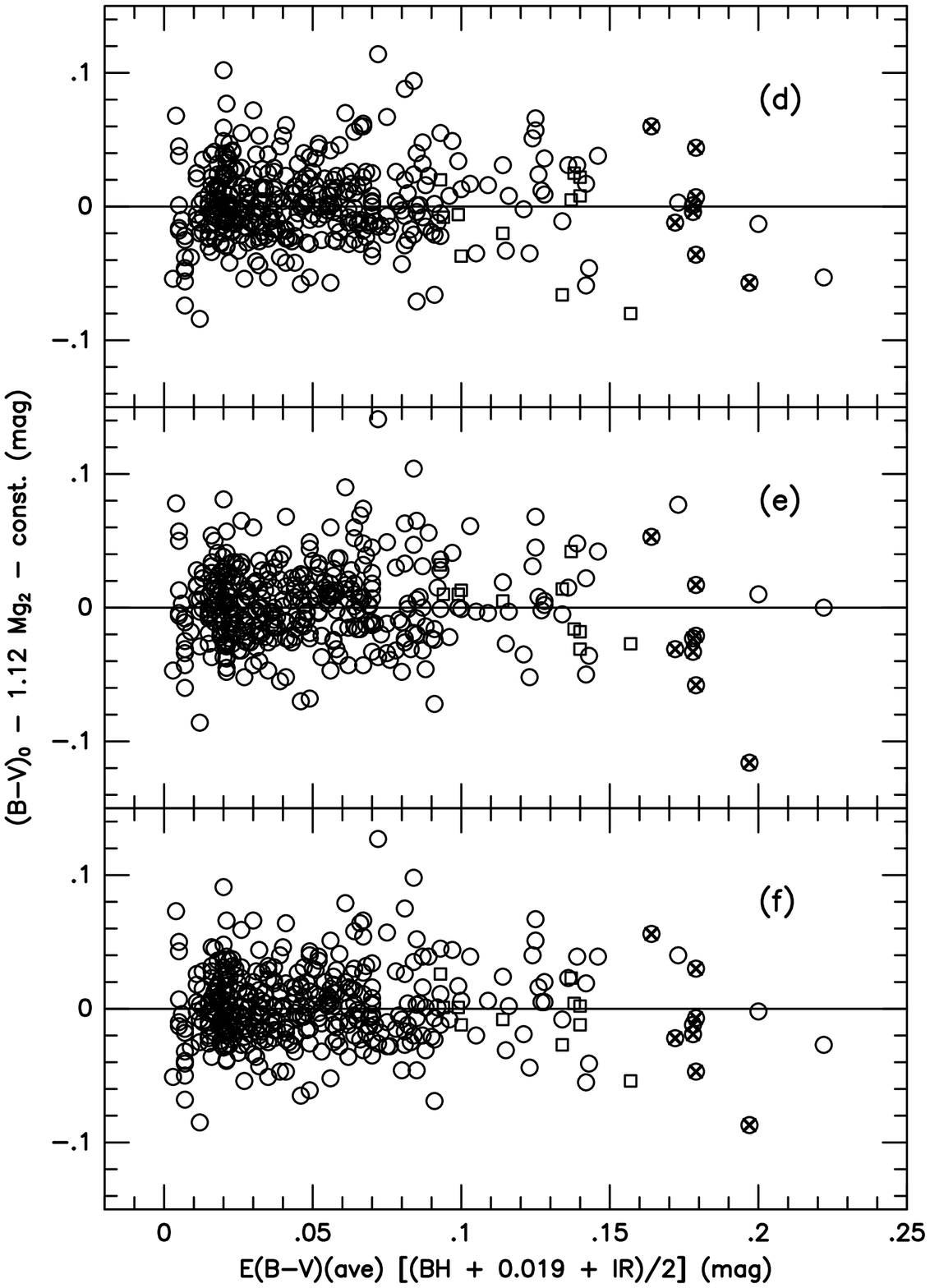}
\figcaption{(continued): (d) for the straight IR predictions; 
(e) the BH+0.019 mag predictions; and (f) the average of BH+0.019 
and IR predictions. \label{fig2d}}
\end{figure}

\begin{figure}
\epsscale{0.7}
\plotone{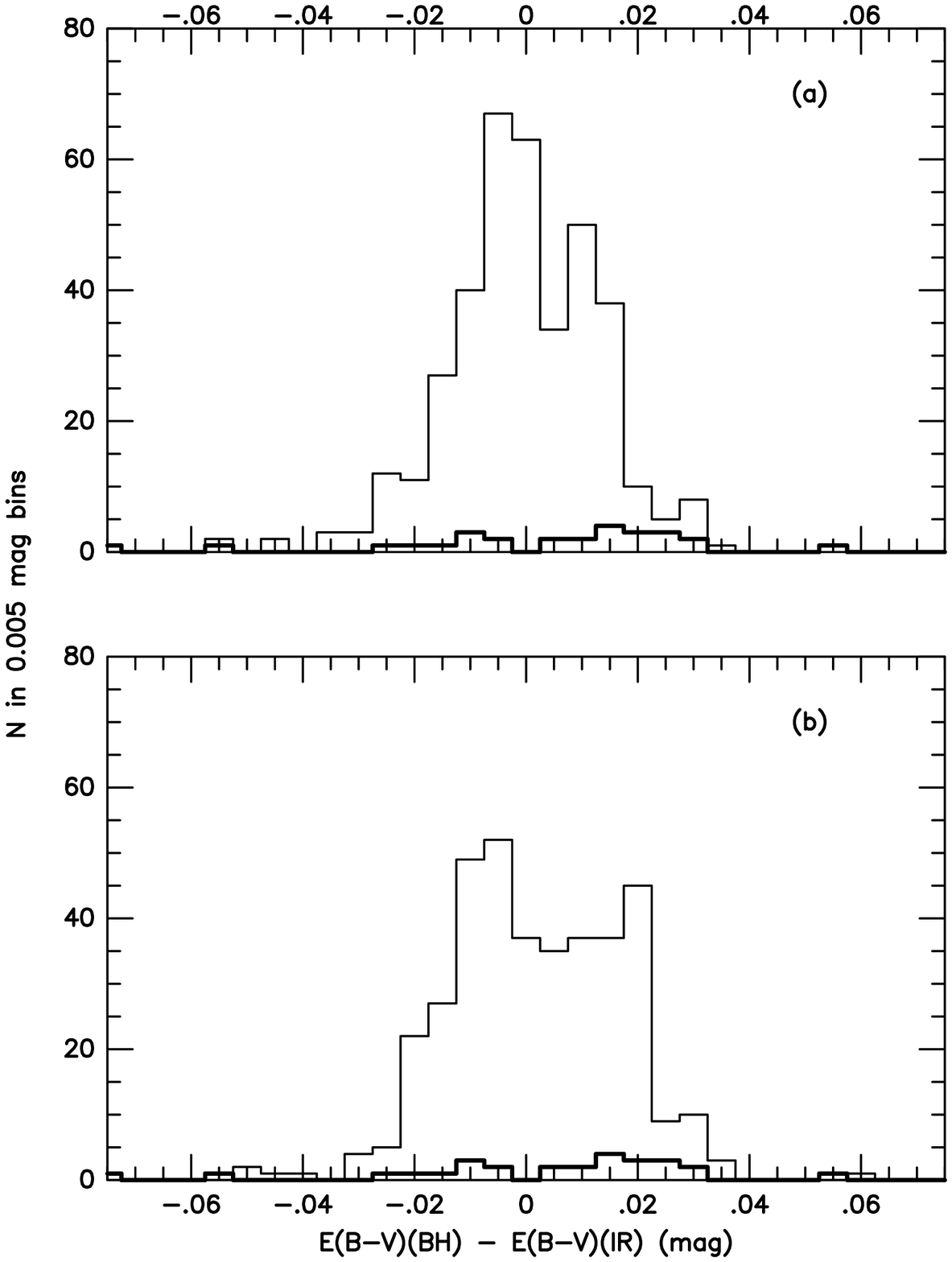}
\figcaption{The histograms of E(B-V) prediction differences, in
the sense BH minus IR, for (a) the BH minus IR-0.016 predictions;
and (b) for the BH+0.019 minus IR predictions. As with Figure~1,
the histograms for the highly-reddened galaxies are given as a dark line.
\label{fig3}}
\end{figure}

\begin{figure}
\epsscale{0.7}
\plotone{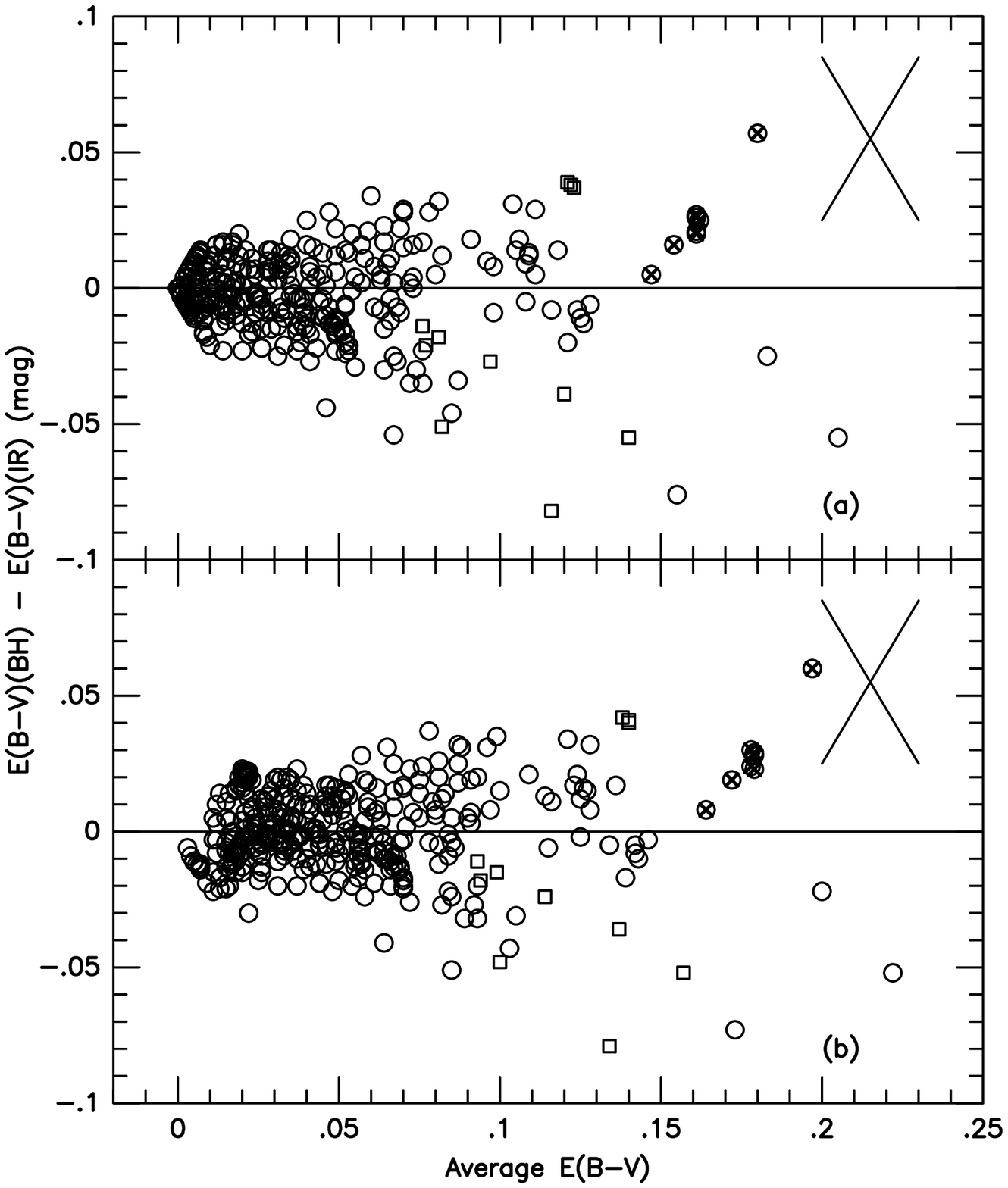}
\figcaption{Graphs of the difference, BH minus IR predictions for
reddening versus the average reddening for the two predictions:
BH, IR-0.016 values; b) BH+0.019, IR values.  If the difference
between the two reddening predictions is systematic, then they
should follow the lines plotted in each graph. The symbols for
the galaxies in the Perseus cluster and in the high gas-to-dust
region in the southern hemisphere region are as in Figure~2. 
\label{fig4}}
\end{figure}

\begin{figure}
\epsscale{0.7}
\plotone{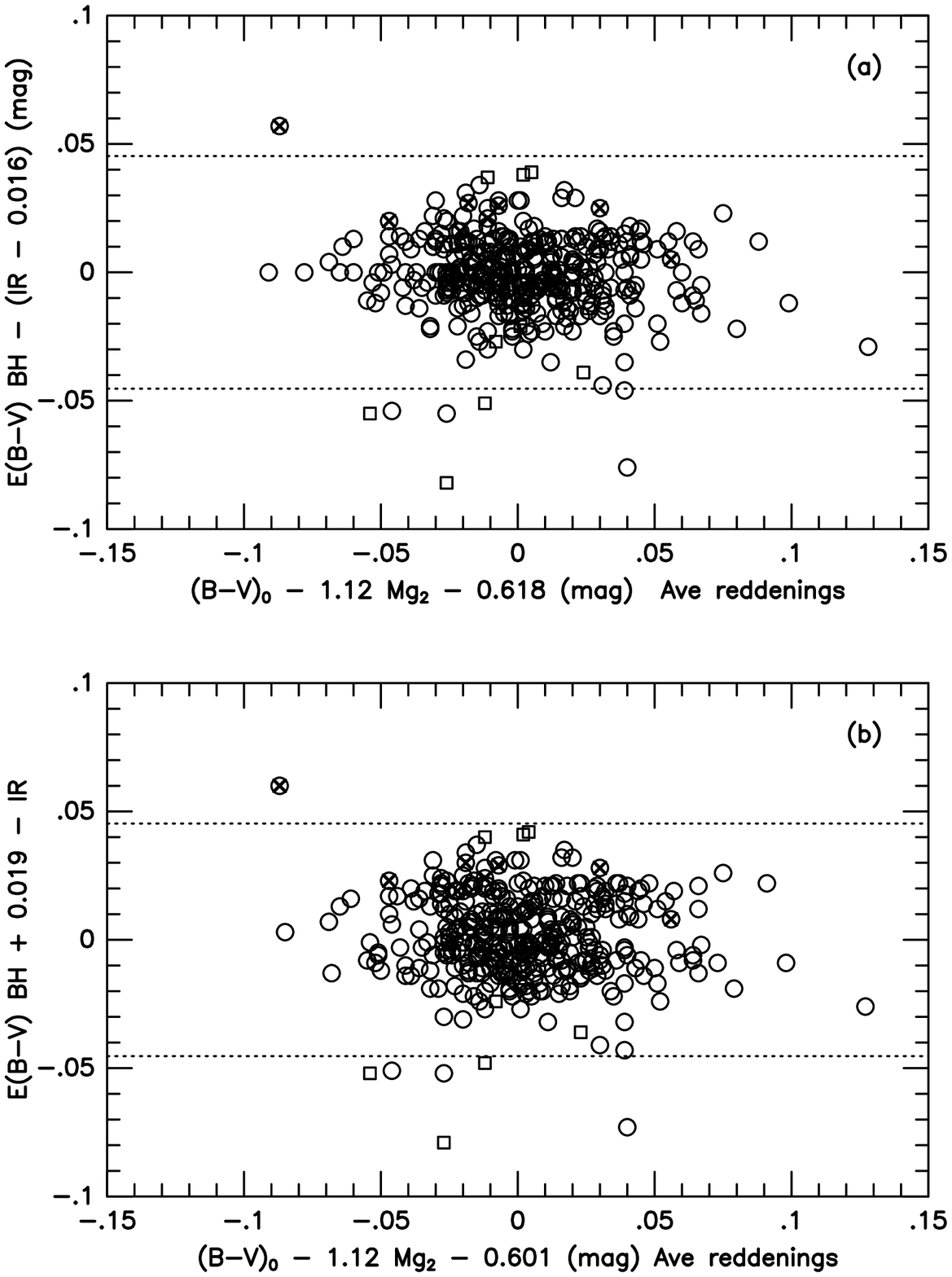}
\figcaption{Graphs of the difference, BH minus IR predictions for
reddening versus the scatter in the (B-V)$_0$-Mg$_2$ relation
using the average reddenings: BH, IR-0.016 values; b) BH+0.019, 
IR values.  The symbols for the galaxies in the Perseus cluster 
and in the high gas-to-dust, southern hemisphere region are as 
in Figure~2. 
\label{fig5}}
\end{figure}


\begin{thebibliography}{}

\bibitem[Appenzeller(1975)]{ap75} Appenzeller, I., 1975, A\&A, 38, 113

\bibitem[Bender, Burstein, \& Faber(1993)]{bbf93} Bender, R., 
Burstein, D., \& Faber, S.M. 1993, \apj, 411, 153

\bibitem[Berdyugin \& Teerikorpi(2002)]{bt02} Berdyugin, A. \&
Teerikorpi, P., 2002, A\&A 384, 1050

\bibitem[Boulanger, et al.(1996)]{betal96} Boulanger, F., Abergel, A.,
Bernard, J.-P., Burton, W.B., Desert, F.-X., Hartmann, D., Lagache,
G., \& Puget, J.-L., 1996, A\&A, 312, 256

\bibitem[Burstein et al.(1987)]{dbetal87} Burstein, D., Davies, R.L., 
Dressler, A., Faber, S.M., Stone, R.P.S., Lynden-Bell, D., Terlevich,
R. \& Wegner, G. 1987, \apjs 64, 601

\bibitem[Burstein et al.(1988)]{dbetal88} Burstein, D., Davies, R.L., 
Dressler, A., Faber, S.M., Lynden-Bell, D., Terlevich, R., \& 
Wegner, G., 1988, Towards Understanding Galaxies at High Redshift, 
ed. R.G. Kron \& A. Renzini, (Kluwer, Dordrecht), p. 17

\bibitem[Burstein \& Heiles(1978a)]{bh78a} Burstein, D. \& Heiles, C.
1978a, Ap Lett, 19, 69

\bibitem[Burstein \& Heiles(1978b)]{bh78b} Burstein, D. \& Heiles, C.
1978b, \apj, 225, 40

\bibitem[Burstein \& Heiles(1982)]{bh82} Burstein, D. \& Heiles, C.
1982, \aj, 87, 1162

\bibitem[Burstein \& Heiles(1984)]{bh84} Burstein, D. \& Heiles, C.
1984, \apjs, 54, 33

\bibitem[Colless et al.(2001)]{mcetal01} Colless, M., Saglia, R.P., 
Burstein, D., Davies, R.L., McMahan, R.K. Jr., \& Wegner, G., 2001, 
\mnras, 321, 277

\bibitem[Davies et et.(1987)]{rldetal87}  Davies, R.L., Burstein, D., 
Dressler, A., Faber, S.M., Lynden-Bell, D., Terlevich, R., 
\& Wegner, G. 1987, \apjs, 64, 581

\bibitem[de Vaucouleurs, de Vaucouleurs, \& Corwin(1976)]{deV76}
de Vaucouleurs, G., de Vaucouleurs, A., \& Corwin, H.G. 1976,
The 2nd Reference Catalogue of Bright Galaxies, (Univ. of Texas
Press, Austin).

\bibitem[Faber et al.(1989)]{smfetal89} Faber, S.M., Wegner, G., 
Burstein, D., Davies, R.L., Dressler, A., Lynden-Bell, D., \& 
Terlevich, R., 1989, \apjs, 69, 763

\bibitem[Greenberg \& Li(1995)]{gl95} Greenberg, J.M. \& Li, A.,
1995, in The Opacity of Spiral Disks, ed. J.I. Davies \& D. Burstein,
(Kluwer, The Netherlands), p. 19

\bibitem[Heiles(1976)]{h76} Heiles, C., 1976, \apj, 204, 379

\bibitem[Heiles(1979)]{ch79} Heiles, C., 1979, \apj, 229, 533

\bibitem[Hudson(1999)]{hudson99} Hudson, M., 1999, \pasp, 111, 57

\bibitem[Li \& Greenberg(1997)]{lg97} Li, A. \& Greenberg, J.M.,
1997, A\&A, 323, 566

\bibitem[Lynden-Bell et al.(1988)]{dlbetal88} Lynden-Bell, D., 
Faber, S.M., Burstein, D., Davies, R.L., Dressler, A., 
Terlevich, R., \& Wegner, G. 1988, \apj, 326, 19

\bibitem[Markannen(1979)]{mark79} Markannen, T., 1979, A\&A, 74, 201

\bibitem[Sandage(1973)]{as73} Sandage, A., 1973, \apj, 183, 711

\bibitem[Sandage(1975)]{as75} Sandage, A., 1975, \apj, 202, 563

\bibitem[Schlegel, Finkbeiner, \& Davis(1998)]{sfd98} Schlegel, D. J.,
Finkbeiner, D. P., \&  Davis, M. 1998, \apj, 500, 525

\bibitem[Strittmatter, et al.(1984)]{sbpo85} Strittmatter, R.E.,
Balasubrahmanyan, V.K., Protheroe, R.J., \& Ormes, J.F., 1985,
A\&A, 143, 249

\bibitem[Teerikorpi(1990)]{t90} Teerikorpi, P., 1990, A\&A, 235, 362

\bibitem[Trager, et al.(2000)]{trager00} Trager, S.C., Faber, S.M.,
Worthey, G., \& Gonzalez, J.J., 2000, \aj, 120, 165

\bibitem[Tully(1988)]{t88} Tully, R.B., 1988, Nearby Galaxies Catalog,
Cambridge Univ Press, New York

\bibitem[VandenBerg, Bolte, \& Stetson(1990)]{vbs90} VandenBerg, D.A.,
Bolte, M., \& Stetson, P.B. 1990, \aj, 100, 445

\end{thebibliography}
\end{document}